\documentclass{kapproc} % Computer Modern font calls
\usepackage{psfig}
\RequirePackage{graphicx}%
\RequirePackage{epsf}%

\upperandlowercase
\let\footnote\savefootnote
\let\footnotetext\savefootnotetext 
\let\footnoterule\savefootnoterule 
\kluwerbib % will produce  bibliography 
\begin{document}

%------------ article title  ------------------->>

\articletitle{Wolf-Rayet Stars As IMF Probes}

%% optional, to supply a subtitle:
%\articlesubtitle{Spineto@50}

%% Supply a shorter version of the title for the running head:
\chaptitlerunninghead{Wolf-Rayet Stars}

%------ author/affiliation choices -------------->>

%% Single author or several authors with same affiliation

 \author{Claus Leitherer}
 \affil{Space Telescope Science Institute, 3700 San Martin Drive, Baltimore, MD 21218, USA}
 \email{leitherer@stsci.edu}

%% Multiple authors, multiple affiliations

%\author{First Author\altaffilmark{1}, Second Author\altaffilmark{2}, 
%         Third Author\altaffilmark{1,3}}

%\affil{\altaffilmark{1}Institute, Address, Country, \\ 
%\altaffilmark{2}Institute, Address, Country, \\
%\altaffilmark{3}Institute, Address, Country}

%\email{author1@add1,author2@add2,author3@add3}

% abstract
 \begin{abstract}
Wolf-Rayet stars are the evolved descendents of massive stars. Their extraordinary properties make them useful tracers of the stellar initial mass function (IMF) in a young stellar population. I discuss how the interpretation of spectral diagnostics are complicated by the interplay of stellar, nebular, and dust properties. There is mounting observational evidence for spatial inhomogeneities in the gas and dust distribution. The interplay of these inhomogeneities can significantly alter frequently used star-formation and IMF indicators. Specific examples presented in this contribution are the starburst galaxies NGC 1614, NGC 2798, and NGC~3125.
 \end{abstract}

%------------ body of article ------------------->>
\section{Background}

Star formation in powerful starbursts is an ubiquitous phenomenon both locally and at the highest observable 
redshifts. A prominent subset of the starburst class are the so-called Wolf-Rayet (\mbox{W-R}) galaxies, whose 
observational characteristic is the broad emission bump around 4640 -- 4690 \AA\ (Conti 1991). The compilation 
by Schaerer et al. (1999) lists 139 members. 

W-R galaxies are important because they permit a study of the star-formation properties via {\em stellar} spectral 
features, as opposed to indirect tracers based on gas and/or dust emission. Hot stars are notoriously elusive even 
in the strongest starbursts because their spectral signatures are too weak, coincide with nebular emission lines, 
or are in the satellite-ultraviolet (UV). W-R stars are the only hot, massive stellar species detectable at optical wavelengths. This is because they have the strongest stellar winds, which in combination with their high temperatures produce broad ($\sim$1000~km~s$^{-1}$) emission lines not coinciding with emission from H~II regions. Examples are N~III $\lambda$4640, C~III $\lambda$4650, He~II $\lambda1640$ and $\lambda$4686, C~III $\lambda$5696, and C~IV $\lambda$5808. The mere detection of such features proves the presence of stars with masses above 40 -- 60~M$_\odot$ (depending on chemical composition) since only stars this massive can evolve into the W-R phase. This powerful diagnostic can be used for, e.g., inferring a massive star population when the space-UV is inaccessible due to dust (infrared [IR]-luminous galaxies), or when broad nebular lines veil the O stars (Seyfert2 galaxies). Vice versa, the relative strength of the various W-R features gives clues on the distribution of W-R subtypes, which in turn provides much sought constraints on stellar evolution theory.

\section{W-R stars as proxies for ionizing O stars}

We are currently involved in an optical+near-IR survey of luminous starburst galaxies (Le\~ao, Leitherer, \& Bresolin, in prep.). Our goals are (i) to establish an unbiased sample of metal-rich W-R galaxies, (ii) to investigate the 
stellar content from purely stellar tracers, (iii) to cross-calibrate stellar and nebular diagnostics in metal-rich 
galaxies, (iv) to correlate the derived stellar initial mass function with host galaxy parameters, and (v) to test 
stellar evolution theory in high-metallicity environments. Our new survey of starburst galaxies is drawn from an IRAS sample of Lehnert \& Heckman (1995). The galaxies have $10 < \log L_{\rm{IR}}/$L$_\odot < 11.5$, warm IR colors, and are not AGN dominated. 
Detection of the W-R features, together with other standard diagnostics allows 
us to probe the stellar content with a suite of models we have developed (Leitherer et al. 1999). In particular, we can search for or against evidence of a peculiar IMF at high metallicity, as indicated, e.g., by IR observations.

Thornley et al. (2000) carried out an ISO spectroscopic
survey of 27 starburst galaxies 
with a range of luminosities from $10^8$ to $10^{12}$~L$_\odot$. 
The [Ne~III] 15.6~$\mu$m and [Ne~II] 12.8~$\mu$m lines are
particularly useful. The ionization
 potentials of neutral and ionized Ne are 22 and 41~eV, respectively,
the two lines are very close in wavelength (12.8 and 15.6~$\mu$m), 
and they have similar critical densities. This makes the line ratio a sensitive
probe for various star formation parameters, in particular the upper mass 
cutoff of the IMF ($M_{\rm{up}}$) and the age and duration of the starburst.
Photoionization models of Thornley et al. and Rigby \& Rieke (2004) suggest
that stars more 
massive than about 35 to 40~M$_\odot$ are deficient in the observed sample. 
Either they never formed because of a peculiar IMF, or they have already 
disappeared due to aging effects. This result echos that obtained
from ground-based near-IR spectroscopy: the strategic recombination lines
are powered by a soft radiation field originating from stars less massive than
$\sim$40~M$_\odot$ (Doyon et al. 1992).
An upper-mass cutoff as low as 40~M$_\odot$, however, is difficult to reconcile with 
the ubiquitous evidence of very massive stars with
masses of up to 100~M$_\odot$ in many starburst regions (e.g., 
Leitherer et al. 1996; Massey \&
Hunter 1998; Gonz\'alez Delgado et al. 2002). 
Therefore the alternative explanation,
an aged starburst seems more plausible. Under this assumption, stars of 
masses 50~--~100~M$_\odot$ are initially formed in most galaxies, but the
starbursts are observed at an epoch when these stars are no longer present.
This implies that the inferred burst durations must be less than a few Myr. Such short burst time scales are surprising,
in particular for luminous, starburst galaxies whose dynamical time scales 
can exceed tens of Myr. Both the peculiar IMF or the short starburst time scales in dusty,
IR-bright starbursts are quite unexpected and pose a challenge to 
conventional models in which the starburst is fed by gas inflow to
the nucleus over tens of Myr as a result of angular momentum loss. 

%% Double captions:
\begin{figure}[ht]
\sidebyside
{\centerline{\psfig{file=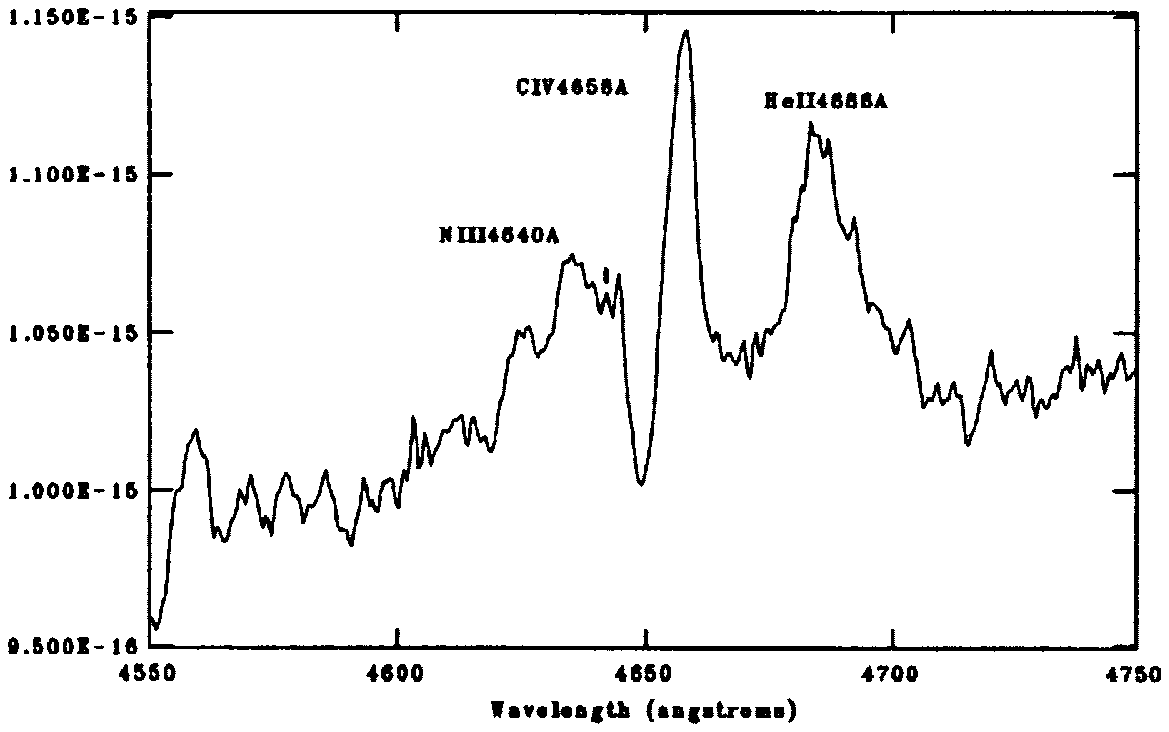,height=1.45in}}
\caption{Close-up view of the spectral region around 4650~\AA\ in the starburst galaxy NGC 1614.}}
{\centerline{\psfig{file=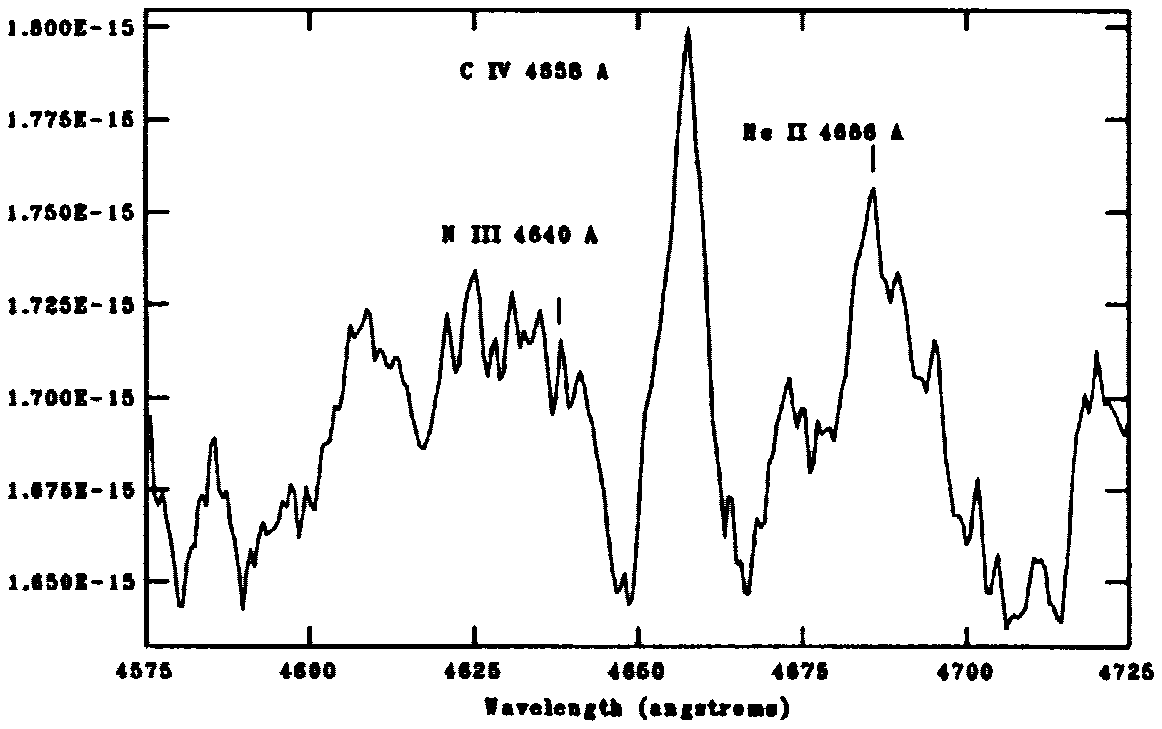,height=1.45in}}
\caption{Same as Fig. 1, but for NGC 2798. Note the broad emission-line features from W-R stars.}}
\end{figure}

Early results from our W-R survey urge caution when relating nebular IMF tracers to the actual stellar content. Several metal-rich starburst galaxies exhibiting a soft radiation field do in fact have a substantial \mbox{W-R} population. As an example, we reproduce in Figs.~1 and 2 portions of Keck LRIS spectra around 4650~\AA\ of NGC~1614 and NGC~2798, two archetypal starburst galaxies with low nebular excitation. We clearly detect the tell-tale signatures of W-R stars in these galaxies. Unless stellar evolution proceeds differently than in our Galaxy, massive O stars must be present as well. The fact that we do not {\em directly} observe these O stars is no contradiction: their spectral lines will be hidden by coinciding nebular emission lines. We should, however, detect their ionizing radiation in a proportion predicted by the measured number of W-R stars and the expected ratio of W-R/O stars. The deficit of radiation suggests that indirect star-formation tracers, like nebular lines, still
require careful calibration, in particular when applied to dusty,
metal-rich starbursts.

After advocating the use of W-R stars to constrain the IMF, we will present a cautionary view point in the next section. Interpretation of the stellar W-R feature may sometimes  be complicated by the inhomogeneous structure of the surrounding interstellar medium (ISM).

\section{The extraordinary W-R cluster in NGC 3125}

As part of a larger project to quantify the stellar and interstellar properties of local galaxies undergoing active star formation (Chandar et al., in prep.), we have obtained HST Space Telescope Imaging Spectrograph (STIS) long-slit far- and near-UV spectra for 15 local starburst galaxies. The target galaxies were selected to cover a broad range of morphologies, chemical composition, and luminosity. Most importantly, target selection was not based on W-R content. 

The UV counterpart of the optical He~II $\lambda$4686 line is the 3 $\rightarrow$ 2 transition of He$^+$ at 1640~\AA.
  Broad He~II $\lambda$1640 emission is seen in the UV spectra of individual Galactic and Magellanic Cloud W-R {\em stars} (e.g., Conti \& Morris 1990) but is not prevalent in the integrated spectra of {\em galaxies} because of the overwhelming light contribution from OB stars in the UV. We observe two different line morphologies of He~II $\lambda$1640 emission in local starburst galaxies. Some galaxies show narrow, nebular He~II emission, while others have a broader profile. If massive stars are forming, He~II $\lambda$1640 can sometimes appear as a nebular recombination emission line (e.g., Garnett et al. 1991), which has a characteristic narrow-line morphology. In the following we will focus on one outstanding starburst region in our sample: NGC~3125-1.  The measured He~II FWHM of $5.9 \pm 0.4$~\AA\ is clearly broader than the estimated instrumental profile of 2.6~\AA. The corrected width of the He~II $\lambda$1640 line is 1000~km~s$^{-1}$. Similar line widths are typically found in the latest subtypes of individual WN stars.

NGC 3125-1 has by far the largest He~II $\lambda$1640 equivalent width in our sample. In Fig.~3 we show the UV spectrum of this object. The 1640~\AA\ emission is quite prominent and appears significantly stronger than the C~IV emission at 1550~\AA. In addition to this feature, we also see N~IV $\lambda$1488 and N~IV $\lambda$1720 emission typically found in \mbox{W-R} stars, consistent with our interpretation that the strong He~II $\lambda$1640 emission arises in the winds of massive stars. We estimate that 6100 WNL stars reside in this starburst region, making it the most W-R-rich known example of an individual starburst cluster in the local universe. What makes this region extraordinary is the number of W-R stars relative to other massive stars. The W-R features in Fig.~3 are almost undiluted compared with those seen in single W-R stars. Therefore, most of the continuum light must be emitted by the very same W-R stars. Quantitative modeling leads to approximately equal W-R and O-star numbers. Such an extreme ratio is excluded by stellar evolution models, even for a most unusual IMF.

\begin{figure}[ht]
\centerline{\psfig{file=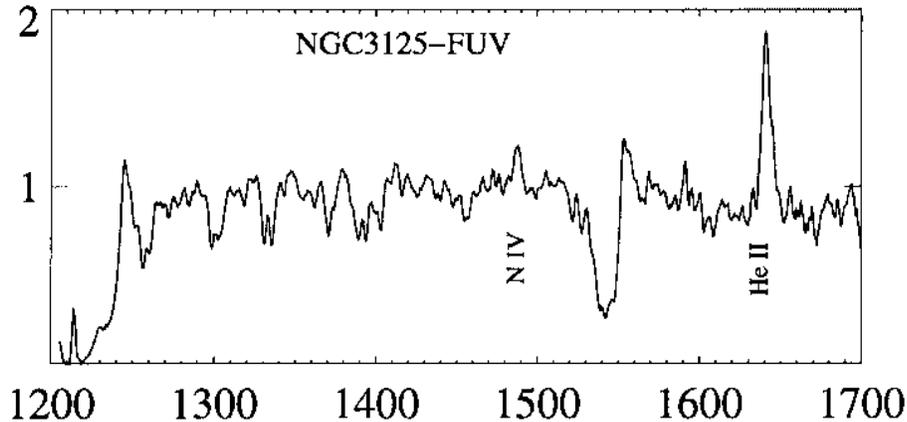,height=2.2in}}
\caption{Far-UV spectral region of the starburst region NGC 3125-1. Strong, broad He II at 1640~\AA\ indicates several thousand W-R stars. The W-R stars provide a substantial fraction of the continuum light as well (Chandar et al. 2004).}
\end{figure}

  How can we reconcile these observations with our current understanding of W-R stars? The UV continuum slope is normally dominated by OB stars; however, in this case the large number of W-R stars also makes a significant contribution. One possible explanation is that we are seeing the W-R stars through a ``hole'' where their energetic winds have blown out the natal cocoon earlier than have the OB stars. The often large reddening derived in starburst regions is consistent with the generally accepted scenario in which very young clusters remain embedded in their natal material until energetic stellar winds from evolving massive stars blow out the surrounding gas and dust. 

     If W-R stars are preferentially less attenuated than OB stars in NGC 3125-1, the {\em equivalent width} of He~II $\lambda$1640 and other W-R lines is skewed towards artificially large values because for a standard stellar population surrounded by a homogeneous ISM, the continuum is emitted by the less massive OB stars. If the ISM is inhomogeneous, a textbook property of a spectral line associated with a single star changes: the equivalent width becomes reddening dependent.

\section{Conclusions}

Determining the massive-star IMF in starburst regions often relies on indirect methods, such as the recombination luminosity of the ionized gas. While this technique has been demonstrated to be quite reliable in normal galaxies (Kewley et al. 2002), it may break down in extreme cases. The conditions prevailing in dusty, metal-rich starbursts may lead to a decoupling of the emitted ionizing and the observed recombination luminosity. The reasons are still not fully understood but may be related to a combination of the destruction of photons by dust and the spatial morphology of the dust. The latter effect is particularly relevant in powerful starbursts where supernovae and stellar winds can have large effects on the ISM. Starburst models as well as observations of galactic superwinds indicate that 1\% to 2\% of the bolometric luminosity of starbursts 
is converted into mechanical luminosity. This mechanical energy input can remove interstellar gas from 
the star-formation site on a time scale much shorter than the gas consumption time scale, which can be hundreds of Myr. The resulting ISM structure invalidates the often made assumption of isotropy and homogeneity. As a result, IMF indicators can mimic anomalous stellar populations. In some cases, even purely stellar indicators may change. If different stellar phases are associated with different dust columns, the stellar equivalent widths in the integrated population spectrum may probe dust morphologies, rather than the IMF. 

\begin{acknowledgments}
I am grateful to Rupali Chandar and Jo\~ao Le\~ao for several comments which improved the manuscript.
\end{acknowledgments}

\begin{chapthebibliography}{}
%\bibitem is optional
Chandar, R., Leitherer, C., \& Tremonti, C. A. 2004, ApJ, 604, 153

Conti, P. 1991, ApJ, 377, 115

Conti, P., \& Morris, P. W. 1990, AJ, 99, 898

Doyon, R., Puxley, P. J., \& Joseph, R. D. 1992, ApJ, 397, 117 

Garnett, D. R., Kennicutt, R. C., Jr., Chu, Y.-H., \& Skillman, E. D. 1991, PASP, 103, 850

Gonz\'alez Delgado, R. M., Leitherer, C., Stasi\'nska, G., \& Heckman, T. M. 2002,
ApJ, 580, 824

Kewley, L.~J., Geller, M.~J., Jansen, R.~A., \& Dopita, M.~A. 2002, AJ, 
124, 3135 

Lehnert, M. D., \& Heckman, T. M. 1995, ApJS, 97, 89 

Leitherer, C., et al. 1999, ApJS, 123, 3

Leitherer, C., Vacca, W. D., Conti, P. S., Filippenko, A. V., 
    Robert, C., \& Sargent, W. L. W. 1996, ApJ, 465, 717

Massey, P., \& Hunter, D. A. 1998, ApJ, 493, 180

Rigby, J. R., \& Rieke, G. H. 2004, ApJ, 606, 237 

Schaerer, D., Contini, T., \& Pindao, M.\ 1999, A\&AS, 136, 35 

Thornley, M.~D., Schreiber, N.~M.~F., Lutz, D., Genzel, R., Spoon, H.~W.~W., Kunze, D., 
     \& Sternberg, A. 2000, ApJ, 539, 641 

\end{chapthebibliography}

\end{document}